\documentclass[a4paper,11pt]{article}
\usepackage{graphicx}
\usepackage{color}
\usepackage{amsmath}
\usepackage{amssymb}
\usepackage{graphicx}
\usepackage{dcolumn}
\usepackage{bm}
\usepackage{color}
\usepackage{enumitem}

\begin{document}

\begin {center}

{\Large {\textsc{A linear nonequilibrium thermodynamics approach to \\optimization of thermoelectric devices}}}

\end{center}

\vspace{1.5cm}

\noindent Henni Ouerdane\\
Laboratoire CRISMAT, UMR 6508 CNRS, ENSICAEN et Universit\'e de Caen Basse Normandie, 6 Boulevard Mar\'echal Juin, F-14050 Caen, France, ~ and ~ Universit\'e Paris Diderot, Sorbonne Paris Cit\'e, Institut des Energies de Demain (IED), 75205 Paris, France\\

\noindent Christophe Goupil\\
Laboratoire CRISMAT, UMR 6508 CNRS, ENSICAEN et Universit\'e de Caen Basse Normandie, 6 Boulevard Mar\'echal Juin, F-14050 Caen, France, ~ and ~ Universit\'e Paris Diderot, Sorbonne Paris Cit\'e, Institut des Energies de Demain (IED), 75205 Paris, France\\

\noindent Yann Apertet\\
Institut d'Electronique Fondamentale, Universit\'e Paris-Sud, CNRS, UMR 8622, F-91405 Orsay, France\\

\noindent Aur\'elie Michot\\
CNRT Mat\'eriaux UMS CNRS 3318, 6 Boulevard Mar\'echal Juin, F-14050 Caen Cedex, France\\

\noindent Adel Abbout\\
Laboratoire CRISMAT, UMR 6508 CNRS, ENSICAEN et Universit\'e de Caen Basse Normandie, 6 Boulevard Mar\'echal Juin, F-14050 Caen, France

\vspace{2.0cm}

\section{Introduction}
At the standard macroscopic scale various technologies exist for waste energy harvesting and conversion: heat exchangers for energy storage and routing, heat pumps, organic Rankine cycle, and thermoelectricity. Depending on the specific working conditions, one technology may be viewed as more efficient than another; for instance, thermoelectricity is more appropriate in cases where the temperature difference between the heat source and sink is not too large. Efforts are invested in the improvement of thermoelectric devices in terms of properties and range of applications because their conversion efficiency is not size-dependent and the typical device does not contain moving parts. These qualities, of paramount importance in view of applications at the mesoscale in the microelectronics industry, recently provided a new impetus for research in the field of thermoelectricity. Tremendous progress in the understanding and mastering of thermoelectric systems has been made since the pioneering works of Seebeck \cite{Seebeck} and Peltier \cite{Peltier}, but much remains to be done in order to improve the energy conversion efficiency at maximum output power. Indeed, even at the macroscale the best energy conversion efficiency of thermoelectric devices typically are of the order of 10 \% of the efficiency of the ideal Carnot thermodynamic cycle.

A ``good'' thermoelectric material has a large thermopower and a high electrical conductivity to thermal conductivity ratio. The properties of a given material may usually be qualified as good in a very limited range of temperatures though. Recent advances in the physics and engineering of semiconductors and strongly correlated materials have permitted great progress by way of optimization of the materials' characteristics, thus offering interesting prospects for device performance and range of operation. But, with practical purposes in mind, one must consider that a real device is not a perfect theoretical system, and that its thermal contacts, through heat exchangers, with the temperature reservoirs are usually far from ideal too. A poor device design and neglect of the quality of the thermal contacts can only yield poor device performance however good is the thermoelectric material. This clearly means that there are a number of truly important technological challenges which must be met; high quality brazing is one of them. From a theoretical/modeling viewpoint, it is also necessary to develop models that capture the essential characteristics of thermoelectric devices operating in realistic working conditions. Indeed, to make the best possible use of the best materials available, one needs to understand how the internal laws of a device may be appropriately associated to the laws that govern its interaction with its environment. We propose here a reflection along these lines.

The link between the intrinsic properties of a thermoelectric device and its performance usually is given by the so-called figure of merit. While much work is devoted to find means to increase the figure of merit, our present work rather focuses on how to achieve optimal working conditions. Thermoelectric devices can be described as heat engines connected to two temperature reservoirs. In these systems, transport of heat and transport of electric charges are strongly coupled. Not too far from equilibrium these transport phenomena obey linear phenomenological laws such as those given in Table~\ref{tab:table1}; so a general macroscopic description of thermoelectric systems is in essence phenomenological. Linear nonequilibrium thermodynamics provides a most convenient framework to characterize the device properties and the working conditions to achieve various operation modes. As we shall see in this chapter the approximations we make, which are well controlled, are used to obtain analytical expressions that facilitate the analysis, discussion and understanding of the physical concepts we study.

\begin{table*}
\caption{\label{tab:table1}Examples of linear phenomenological laws.}
\begin{tabular}{|c|c|c|}\hline\hline
{\bf variables}&{\bf transport coefficient}&{\bf expression} and {\bf name} \\
\hline
\rule{0pt}{5ex}electrical current density and electric field&electrical conductivity&$\bm J = \sigma \bm E \equiv - \sigma {\bm \nabla} \varphi$\\
~&~& Ohm's law\\
\rule{0pt}{5ex}particle flux and density&diffusion coefficient&$\bm J_{N} = -D {\bm \nabla} n$\\
~&~&Fick's law\\
\rule{0pt}{5ex}energy flux and temperature&thermal conductivity&$\bm J_{E} = -\kappa {\bm \nabla} T$\\
~&~&Fourier's law\\
\hline\hline
\end{tabular}
\end{table*}

Our approach, presented here for a model of a thermoelectric generator non-ideally coupled to the temperature reservoirs through finite-conductance heat exchangers, is quite an appropriate starting point for an extension down to mesoscopic scale. Assuming a simple resistive load and introducing an effective thermal conductance for the device, we show first that, in addition to electrical impedance matching, conditions that permit thermal impedance matching must be also satisfied in order to achieve optimal device performance. The problem of efficiency at maximum power is central in our work, but it becomes quite tricky as soon as it is addressed at a fundamental level. It is fortunate that the model system we study allows derivation of simple formulas at the cost of approximations that are perfectly reasonable in the framework of linear response.

The thermodynamic formulation that we use is that of Callen\footnote{For convenience, Callen formulates the postulates of thermodynamics only for simple systems, defined as systems that are large enough, macroscopically homogeneous, isotropic and uncharged; the surface effects can be neglected, and no external electric, magnetic, or gravitational fields acts on these systems.} \cite{CallenBook}: in its modern form, due to Callen in 1960, the equilibrium thermodynamics can be summarized by the main postulate of the existence of an entropy maximum. The postulates, which will be detailed in the next section, assume an understanding of: (i) the distinctions between macroscopic and microscopic variables, and between extensive and intensive macroscopic variables; (ii) the concept of a system surrounded by boundaries that restrict, i.e. hold constant, some or all of the extensive variables of the system; (iii) the definitions of internal energy $U$ and work done on a system $W$, and the concept of heat $Q$, defined through the  first law of thermodynamics: $\delta Q ={\rm d}U - \delta W$.

We start this Chapter with a recap of linear nonequilibrium thermodynamics: a brief overview of some of the basics concepts and tools developed by Onsager \cite{onsager1,onsager2} and Callen \cite{callen1} is necessary to set the scenery. A short presentation of thermoelectric effects in absence of magnetic fields will follow. We will see that the force-flux formalism provides a net description of thermoelectric processes \cite{Domenicali}. Then, we will turn to our analysis of device optimization. The Chapter ends with a discussion on efficiency at maximum power, and an outlook on thermoelectricity at the mesoscopic scale. 

\section{Basic notions of linear nonequilibrium thermodynamics}

For the sake of clarity in the next sections of the chapter, it is useful at this stage to start from the basic definitions and notions. Before getting into nonequilibrium physics, it is useful to remind here some notions concerning equilibrium.

A thermodynamic \emph{system} is usually defined as a collection of a great number of objects characterized by fundamental quantities called extensive and intensive variables that describe its macroscopic properties. A thermodynamic \emph{state} is defined by the specification of some macroscopic physical properties of the system (all the properties are not necessary for a study). Therefore \emph{one} physical system may correspond to \emph{many} thermodynamic systems. We will precise these points below, when the postulates of equilibrium statistical mechanics are stated.

To each set of extensive variables associated to a thermodynamic system, there is a counterpart, i.e. a set of intensive variables. The thermodynamic potentials are constructed from these variables. For example, for a gas of noninteracting molecules, one may consider the following extensive variables: entropy $S$, volume $V$, and particle number $N$, and their coupled intensive variables: temperature $T$, pressure $P$, and chemical potential $\mu$. The internal energy, which is a thermodynamic potential, is given by: $U = TS -PV + \mu N$. The thermodynamic equilibrium is obtained when thermal, mechanical, and chemical equilibria are reached. This may be reformulated as follows: an equilibrium state is reached when all the thermodynamic potentials are minimum. This implies that if the thermodynamic system is in equilibrium, all the parts of this system are in equilibrium.

\subsection{Postulates and origin of irreversibilities}
Thermodynamics is useful to describe equilibrium states, but physical processes are rather characterized by irreversibility and nonequilibrium states. A thermodynamic description of equilibrium states may only yield very incomplete information on the actual processes at work, and thus needs to be extended to account for the rates of the physical processes. Irreversible thermodynamics provides links between the measurable quantities, and nonequilibrium statistical physics provides the tools to compute these.

The framework of nonequilibrium statistical mechanics is essentially rooted in three postulates, two of which concern equilibrium:

\begin{enumerate}
   \item A thermodynamic system, isolated and in equilibrium, is characterized by a very large number of \emph{accessible} microstates; spontaneous transitions occur continually between these microstates.
   \item Each of the accessible microstates has equal a priori probability.
\end{enumerate}

\noindent For a given system, the ensemble of microstates with same energy forms a statistical ensemble called a \emph{microcanonical} ensemble. The a priori probability relates to the principle of indifference: during the course of its evolution, an isolated system in an equilibrium state will experience \emph{all} the accessible microstates at the \emph{same} recurrence rate. Therefore, one may assume that the average of a physical quantity over long times is equal to the average of the same quantity over the microcanonical ensemble. In other words, for a stationary system, there is an equivalence between the performing of many identical measurements on a single system and a single measurement on many replicas of the system. This is the ergodic hypothesis.

The third postulate, in its simplest form reads:

\begin{enumerate}[start=3]
   \item Time symmetry of physical laws: in absence of applied magnetic or Coriolis force fields, their mathematical formulation remains unchanged if the time $t$ is everywhere replaced by $-t$.
\end{enumerate} 

The state of a macrocospic system is defined by its macroscopic parameters, but a macrostate gives no information about the state of an individual component; so if the macroscopic properties of two stationary systems take the same values, these two systems are thermodynamically indistinguishable. The probability that a certain macrostate is realized is determined by the number of microstates that correspond to this macrostate; this number is called the multiplicity of the given macrostate. Since thermodynamic systems are large, macrostates multiplicities are immensely large. Some macrostates are more probable than others so, in a nutshell, irreversibility of processes in macroscopic systems appears as the evolution from the less probable to the more probable configuration in phase space.

Now, why are some macrostates more probable than others? As a matter of fact, irreversibility emerges as a result of different but tightly connected factors \cite{LeBellac}. The first of these is the \emph{very} large number of degrees of freedom in a thermodynamic system. A direct consequence of this large number is that the relation between the probability of the thermodynamic system to be in a macrostate and the occupied phase space volume can only be based on probabilistic arguments. Moreover, the trajectories in phase space are extremely sensitive to the conditions in which the system was initially prepared by application of some constraints; this implies that, after the lifting of some or all of the said constraints, the dynamics that drives the relaxation of the system towards an equilibrium state is by essence chaotic, and hence, in the course of the system evolution, the probability to pass again through the initial macrostate can only decrease and come to be \emph{extremely} small. Though arguments against irreversibility such as Zermelo's paradox based on Poincar\'e's recurrence, and Loschmidt's paradox based on microreversibility were put forward, it is obvious that on the one hand, a Poincar\'e's recurrence time is overwhelmingly large for a thermodynamic system and, on the other hand, there is no such thing as a real, perfectly isolated system: perturbations, as small as they may be but strictly nonzero, are unavoidable.

\subsection{Principle of maximum entropy and time scales}
At the macroscopic scale, the equilibrium states of a system may be characterized by a number of extensive variables $X_i$. If the system is composed of several subsystems, relaxation of some constraints yields changes in the values taken by the variables $X_i$, which correspond to the exchanges between the subsystems. For each equilibrium state, one may define a function $S$ that is positive, continuous and differentiable with respect to the variables $X_i$:

\begin{equation}
S:~ X_i \mapsto S(X_i)
\end{equation}

\noindent The function $S$, called entropy, is extensive: $S$ is the sum of the entropies of the subsystems. The values finally taken by the variables $X_i$ after relaxation of constraints, are those which characterize equilibrium, and which correspond to the maximum of the function $S$.

The extensive variables $X_i$, macroscopic by nature, also differ from the microscopic variables because of the typical time scales over which they evolve: the relaxation time of the microscopic variables being extremely fast, the variables $X_i$ may be qualified as \emph{slow} in comparison. In fact, one may distinguish four well separated time scales:

\begin{enumerate}
   \item The duration of one single collision, $\tau_0$
   \item The collision time, which is the typical time which passes between two consecutive collision events, $\tau_{\rm col}$
   \item The relaxation time towards local equilibrium $\tau_{\rm relax}$
   \item The time necessary for the evolution towards the macroscopic equilibrium, $\tau_{\rm eq}$
\end{enumerate}

These characteristic times satisfy the following inequalities:
\begin{equation}
\tau_0 \ll \tau_{\rm col} \ll \tau_{\rm relax} \ll \tau_{\rm eq}
\end{equation}

Therefore, since the variables $X_i$ are slow, one may define an \emph{instantaneous entropy}, $S(X_i)$, at each step of the relaxation of the variables $X_i$. The differential of the function $S$ is:

\begin{equation}
{\rm d}S = \sum_i\frac{\partial S}{\partial X_i}~{\rm d}X_i  = \sum_i F_i {\rm d}X_i,
\end{equation}

\noindent where each quantity $F_i$ is the intensive variable conjugate of the extensive variable $X_i$.

\subsection{Forces and fluxes}

The notions which follow are best introduced in the case of a discrete system\footnote{One may imagine for instance two separate homogeneous systems initially prepared at two different temperatures and then put in thermal contact through a thin diathermal wall. The thermalization process will trigger a flow of energy from on system to the other.}. Now assume an isolated system composed of two weakly coupled sub-systems to which an extensive variable taking the values $X_i$ and $X_i'$, is associated. One has: $X_i + X_i' = X_i^{(0)} = \mbox{constante}$ and $S(X_i) + S(X_i') = S(X_i^{(0)})$. Then, the equilibrium condition maximizing the total entropy is given by:

\begin{equation}
\frac{\partial S^{(0)}}{\partial X_i}\Big|_{X_i^{(0)}} = \frac{\partial (S+S')}{\partial X_i}~{\rm d}X_i\Big|_{X_i^{(0)}} = 
\frac{\partial S}{\partial X_i} - \frac{\partial S'}{\partial X_i'} = F_i -F_i' = 0
\end{equation}

\noindent The equation above tells us that if the difference ${\mathcal F}_i = F_i -F_i'$ is zero, the system is in equilibrium; otherwise an irreversible process takes place and drives the system to equilibrium. The quantity ${\mathcal F}_i$ thus acts as a \emph{generalized force} (or affinity) allowing the evolution of the system towards equilibrium. In addition, we introduce the rate of variation of the extensive variable $X_i$, which characterizes the response of the system to the applied force:

\begin{equation}
J_i = \frac{{\rm d}X_i}{{\rm d}t}
\end{equation}

\noindent And we see that a given flux cancels if its conjugate affinity cancels. Conversely, a non-zero affinity yields a non-zero conjugated flux. In other words, the relationship between affinities and fluxes characterizes the changes due to irreversible processes.

\subsection{Entropy production and local equilibrium}

For a given out-of-equilibrium system, it is useful to study the rate of variation of the total entropy in order to determine the appropriate forces and fluxes. Retaining the same notation as above, we have:

\begin{equation}
\frac{{\rm d}S^{(0)}}{{\rm d}t} = \sum_i \frac{\partial S^{(0)}}{\partial X_i}~\frac{{\rm d}X_i}{{\rm d}t}
\end{equation}

\noindent This rate of variation also is called \emph{entropy production}. It can be rewritten as:

\begin{equation}
\frac{{\rm d}S^{(0)}}{{\rm d}t} = \sum_i {\mathcal F}_i J_i
\end{equation}

\noindent This rate exhibits a bilinear structure: it is the sum of the products of each flux by its conjugate affinity. Note that this property can be generalized to continuous media.

Large systems may be treated as continuous media, which are assumed to be in equilibrium \emph{locally}. More precisely, a system may be divided in cells of intermediate size, i.e. small enough so that the variables vary only little, but large enough to be considered as thermodynamical sub-systems in contact with their environment. It is then possible to define local thermodynamical quantities that are uniform within each separate cell, but different from one cell to another. With the assumption of local equilibrium, the local entropy $\sigma(\bm r)$, as a function of local thermodynamical quantities, has the same form as the entropy $S$ and the local intensive variables are defined as functional derivatives of $S$. Note that the local equilibrium cells are open to energy and matter transport

\subsection{Entropy balance and miminum entropy theorem}
Entropy is an extensive quantity that is not conserved. The global balance for entropy reads:

\begin{equation}
\frac{{\rm d}S}{{\rm d}t} = - \int_{\Sigma} {\bm J}_S \cdot {\bf n} ~{\rm d}\Sigma + \int_V \sigma_S {\rm d}^3{\bm r} \equiv \frac{{\rm d}S_{\rm exch}}{{\rm d}t} + \frac{{\rm d}S_{\rm int}}{{\rm d}t},
\end{equation}

\noindent where ${\rm d}S_{\rm exch}/{\rm d}t$ is the contribution to ${\rm d}S/{\rm d}t$ due to entropy exchange between the system and its environment (thermostat) and ${\rm d}S_{\rm int}/{\rm d}t$ is the entropy production related to the internal changes of the system. The quantities ${\bm J}_S$ and $\sigma_S$ are the entropy flux and the entropy source respectively. The entropy production characterizes the rate of variation of the entropy of the \emph{global} system: \{system;environment\}. Irreversible phenomena that contribute to an entropy production are called \emph{dissipative} phenomena.

In some circumstances, non-equilibrium states are \emph{steady} states in the sense that they are characterized by state variables $X_i$ which are time-independent. In this case, we can write:

\begin{equation}
\frac{{\rm d}S}{{\rm d}t} = \frac{{\rm d}S_{\rm exch}}{{\rm d}t} + \frac{{\rm d}S_{\rm int}}{{\rm d}t} = 0
\end{equation}

\noindent and, since 

\begin{equation}
\frac{{\rm d}S_{\rm int}}{{\rm d}t} = \int_V \sigma_S {\rm d}^3{\bm r} \geq 0
\end{equation}

\noindent then ${\rm d}S_{\rm exch}/{\rm d}t \le 0$. This implies that in order to maintain a system in a non-equilibrium steady state, entropy must \emph{continually} be transferred from this system to its environment. As shown by Prigogine in 1945, the non-equilibrium steady states correspond to a minimum of entropy production since in a time-dependent system, the rate of entropy production can only decrease monotonically as the system approaches equilibrium \cite{Prigogine}.

\subsection{Linear response and reciprocal relations}
In a continuous medium in local equilibrium, the fluxes depend on their conjugate affinity (direct effect), but \emph{also} on the other affinities (indirect effect). At a given point in space and time $({\bm r}, t)$, the flux $J_i$ can be mathematically defined as dependent on the force ${\mathcal F}_i$, but also on the other forces ${\mathcal F}_{j\neq i}$:

\begin{equation}
J_i({\bm r},t) \equiv J_i({\mathcal F}_1,{\mathcal F}_2,\ldots)
\end{equation}

\noindent Close to equilibrium, $J_i({\bm r},t)$ can be written as a Taylor expansion:

\begin{equation}
J_k({\bm r},t) = \sum_j \frac{\partial J_k}{\partial {\mathcal F}_j}~{\mathcal F}_j
+ \frac{1}{2!} \sum_{i,j} \frac{\partial^2 J_k}{\partial {\mathcal F}_i{\mathcal F}_j}~{\mathcal F}_i{\mathcal F}_j + \ldots
= \sum_k L_{jk}{\mathcal F}_k + \frac{1}{2}\sum_{i,j} L_{ijk}{\mathcal F}_i{\mathcal F}_j + \ldots
\end{equation}

\noindent The quantities $L_{jk}$ are the \emph{first-order} kinetic coefficients; they are given by the equilibrium values of the intensive variables $F_i$. The matrix $\left[{\mathcal L}\right]$ of the kinetic coefficients characterizes the \emph{linear response} of the system.

In the linear regime, the source of entropy reads:

\begin{equation}
\sigma_S = \sum_{i,k}L_{ik}{\mathcal F}_i{\mathcal F}_k
\end{equation}

\noindent Since $\sigma_S \ge 0$, the kinetic coefficients satisfy

\begin{equation}
L_{ii} \ge 0 ~~~\mbox{and} ~~~ L_{ii}L_{kk} \ge \frac{1}{4}\left(L_{ik} + L_{ki}\right)
\end{equation}

\noindent If some processes induce rapid variations of the affinities in space and time, there cannot be any local equilibrium. The kinetic coefficients of linear theory acquire a non-local and a retarded character: the fluxes at a given point and a given time, depend on the affinities at other points in space, and at previous times.

In 1931, Onsager put forward the idea that there exist symmetry and antisymmetry relations between kinetic coefficients \cite{onsager1,onsager2}: the so-called \emph{reciprocal relations} must exist in all thermodynamic systems for which transport and relaxation phenomena are well described by linear laws. The main results can be summarized as follows \cite{Pottier}:\\

\begin{itemize}
   \item Onsager's relation: $L_{ik} = L_{ki}$\\
   
   \item Onsager-Casimir relation: $L_{ik} = \epsilon_i\epsilon_k L_{ki}$\\
   
   \item generalized relations: $L_{ik}({\bm H},{\bm \Omega}) = \epsilon_i\epsilon_kL_{ki}(-{\bm H},-{\bm \Omega})$\\
\end{itemize}

\noindent where ${\bm H}$ and ${\bm \Omega}$ respectively denote a magnetic field and an angular velocity associated to a Coriolis field; the parameters $\epsilon_i$ denote the parity with respect to time reversal: if the quantity studied is invariant under time reversal transformation, it has parity $+1$; otherwise this quantity changes sign, and it has parity $-1$. Onsager's reciprocal relations are rooted in the reversibility of the microscopic equations of motion. 

The response of a system upon which constraints are applied, is the generation of fluxes which correspond to transport phenomena. When the constraints are lifted, relaxation processes drive the system to an equilibrium state. Energy dissipation and entropy production are associated to transport and relaxation processes.

\section{\label{frceflx}Forces and fluxes in thermoelectric systems}

\subsection{Thermoelectric effects}

A naive definition would state that thermoelectricity results from the coupling of Ohm's law and Fourier's law. The thermoelectric effect in a system may rather be viewed as the result of the mutual interference of two irreversible processes occurring simultaneously in this system, namely heat transport and charge carrier transport. In thermoelectricity, three effects are usually described:

\begin{enumerate}
   \item The Seebeck effect, which is the rise of an electromotive force in a thermocouple, i.e. across a dipole composed of two conductors forming two junctions maintained at different temperatures, under zero electric current.
   
   \item The Peltier effect, which is a thermal effect (absorption or production of heat) at the junction of two conductors maintained at the same temperature.  
   
   \item The Thomson effect, which is a thermal effect that goes together with the steady electrical current that flows through a  resistive dipole because of the existence of a temperature gradient applied to the dipole.
\end{enumerate}

\noindent It is important to realize here that these three ``effects'' all boil down to the same process: At the microscopic level, an applied temperature gradient causes the charges to diffuse\footnote{One may see an analogy with a classical gas expansion.}, so the Seebeck, Peltier and Thomson effects are essentially the same phenomenon, i.e. thermoelectricity, which manifests itself differently as the conditions for its observation vary. Broadly speaking, when a temperature difference is imposed across a thermoelectric device, it generates a voltage, and when a voltage is imposed across a thermoelectric device, it generates a temperature difference. The thermoelectric devices can be used to generate electricity, measure temperature, cool or heat objects. For a thermocouple composed of two different materials A and B, the voltage is given by:

\begin{equation}
V_{\rm AB} = \int_{T_1}^{T_2} (\alpha_{\rm B} - \alpha_{\rm A}){\rm d}T,
\end{equation}

\noindent where the parameters $\alpha_{\rm A/B}$ are the Seebeck coefficients or thermopowers.

\subsection{The Onsager-Callen model}

The main assumption of Onsager's work is based on the hypothesis that the system evolution is driven by a minimal production of entropy where each fluctuation of any intensive variable undergoes a restoring force to equilibrium \cite{Rocard1967}. This permits the use of a stationary description with a clear definition of all the thermodynamical potentials, though the system itself produces dissipation. From a thermodynamic point of view this is no more than a definition of a quasi-static process since the system is considered to back to local equilibrium at each time. This leads to the very important result that the classical quasi-static relation between heat and entropy variation ${\rm d} S = \delta Q_{\rm qs}/T$ may be extended to finite time response thermodynamics in the following flux form:

\begin{equation}\label{Entropy Flux}
{\bm J}_{S}=\frac{{\bm J}_{Q}}{T}
\end{equation}

\noindent which allows a continuous thermodynamical description of the system: the thermodynamical equilibrium, with all average fluxes equal to zero, just becomes one possible thermodynamical state for the system.

The domain of validity of Onsager's description is thus limited to processes where entropy production is always minimal. By minimal we do not mean that the system will always take the overall minimal entropy production state, but only the minimal entropy production with respect to the external applied constraints, which are called \emph{working conditions}. These may be fulfilled or not, leading to an overall minimal entropy production that can be very far from its optimal value. Finally one may notice that Onsager's description is no more than a generalization of the fluctuation-dissipation theorem, which assumes that the linear response of a system in a stationary state, and the noisy response of this system are related through the same underlying mechanisms\cite{Pottier,Rocard1967,Callen2,Kubo}.

\subsection{Coupled fluxes}

The Onsager force-flux derivation is obtained from the consideration of the laws of conservation of energy and matter. The expression of the relation between the energy flux ${\bm J}_{E}$, the heat flux ${\bm J}_{Q} $, and the particle flux ${\bm J}_{N}$:

\begin{equation}\label{Energy Flux}
{\bm J}_{E}={\bm J}_{Q}+\mu_{\rm e}{\bm J}_{N}
\end{equation}

\noindent is established by application of the first principle of thermodynamics. Each of these fluxes is the conjugate variable of its thermodynamic potential gradients. In the case of an electron gas, the correct potentials for energy and particles are respectively $1/T$ and $\mu _{\rm e}/T$, and the corresponding forces are: ${\bm F}_{N}={\bm \nabla }(-\mu_{\rm e}/T)$ and ${\bm F}_{E}={\bm \nabla }(1/T)$. Then the linear coupling between forces and fluxes may simply be described by a linear set of coupled equations involving the so-called kinetic coefficient matrix $\left[{\mathcal L}\right] $:

\begin{equation}
\left[ 
\begin{array}{c}
{\bm J}_{N} \\ 
{\bm J}_{E}
\end{array}
\right] =\left[ 
\begin{array}{cc}
L_{NN} & L_{NE} \\ 
L_{EN} & L_{EE}
\end{array}
\right] \left[ 
\begin{array}{c}
{\bm \nabla }(-\frac{\mu_{\rm e}}{T}) \\ 
{\bm \nabla }(\frac{1}{T})
\end{array}
\right]
\end{equation}

\noindent where $L_{NE}=L_{EN}$.

\label{Symmetry of the coefficients}

The symmetry of the off-diagonal terms is a fundamental aspect of Onsager's analysis, since it is equivalent to a minimal entropy production of the system in out-of-equilibrium conditions. It should be noticed that the minimal entropy production is not a general property of out-of-equilibrium processes at all so that Onsager's assumption should be present inside the kinetic matrix $\left[{\mathcal L}\right]$. It is known from linear response theory that linear response and fluctuations inside a dissipative system are closely linked. Then each of the fluctuating potentials experiences a restoring force derived from the others in a symmetric form. From a purely thermodynamic point of view this coincides with the Lechatelier-Braun principle. The equality $L_{NE}=L_{EN}$ is nothing else but the manifestation of the intrinsic symmetry of the coupled fluctuations process. From a microscopic point of view this equality also implies the time reversal symmetry of the processes\footnote{This time reversal symmetry is broken under the application of Coriolis or magnetic forces.}. By extension, at the microscopic scale processes should be ``microreversible'', and ``irreversible thermodynamics'' becomes ``reversible dynamics''.

\subsection{Energy flux and heat flux}

To treat properly heat and electrical currents it is more convenient to consider ${\bm J}_{Q}$ instead of ${\bm J}_{E}$ as the pertinent quantity to be analyzed. Using ${\bm J}_{E}={\bm J}_{Q}+\mu _{\rm e}{\bm J}_{N}$ we obtain:

\begin{equation}
\left[ 
\begin{array}{c}
{\bm J}_{N} \\ 
{\bm J}_{Q}
\end{array}
\right] =\left[ 
\begin{array}{cc}
L_{11} & L_{12} \\ 
L_{21} & L_{22}
\end{array}
\right] \left[ 
\begin{array}{c}
-\frac{1}{T}{\bm \nabla }(\mu _{\rm e}) \\ 
{\bm \nabla }(\frac{1}{T})
\end{array}
\right]
\end{equation}

\noindent with $L_{12}=L_{21}$. Since ${\bm \nabla}(-\mu _{\rm e}/T)=-\mu_{\rm e}{\bm \nabla}(1/T)-1/T{\bm \nabla}(\mu _{\rm e})$ the heat and electronic currents read:

\begin{equation}\label{Onsager}
\left[ 
\begin{array}{c}
{\bm J}_{N} \\ 
{\bm J}_{Q}
\end{array}
\right] =\left[ 
\begin{array}{cc}
L_{NN} & L_{NE}-\mu_{\rm e}L_{NN} \\ 
L_{NE}-\mu_{\rm e}L_{NN} & -2L_{NE}\mu_{\rm e}+L_{EE}+\mu_{\rm e}^{2}L_{NN}
\end{array}
\right] \left[ 
\begin{array}{c}
{\bm \nabla }(-\frac{\mu_{\rm e}}{T}) \\ 
{\bm \nabla }(\frac{1}{T})
\end{array}
\right]
\end{equation}

\noindent with the following relationship between kinetic coefficients:

\begin{eqnarray}
L_{11}&=&L_{NN}\\
L_{12}&=&L_{NE}-\mu_{\rm e}L_{NN}\\
L_{22}&=&L_{EE}-2\mu_{\rm e}L_{EN}+\mu_{\rm e}^{2}L_{NN}
\end{eqnarray}

Note that since the electric field derives from the electrochemical potential we also obtain

\begin{equation}
{\bm E}=-\frac{{\bm \nabla}(\mu_{\rm e})}{e}
\end{equation}

\section{Thermoelectric coefficients}

The thermoelectric coefficients can be derived from the expressions of the electronic and heat flux densities depending on the applied thermodynamic constraints: isothermal, adiabatic, electrically open or closed circuit conditions.

\subsection{Decoupled processes}

Under isothermal conditions the electrical current flux may be written in the form,

\begin{equation}\label{Ohm Law}
{\bm J}_{N}=\frac{-L_{11}}{T}{\bm \nabla }(\mu_{\rm e})
\end{equation}

\noindent This is an expression of the law of Ohm since with ${\bm J}=e{\bm J}_{N}$ we obtain the following relationship between the electrical current density and the electric field:

\begin{equation}
e{\bm J}_{N}={\bm J}=e\frac{-L_{11}}{T}{\bm \nabla }(\mu_{\rm e})=\sigma _{T}\left( -\frac{{\bm \nabla }(\mu_{\rm e})}{e}\right) =\sigma _{T}{\bm E},
\end{equation}

\noindent which contains a definition for the isothermal electrical conductivity expressed as follows:

\begin{equation}\label{electrical conductivity}
\sigma_{T}=\frac{e^{2}}{T}L_{11}
\end{equation}

Now, if we consider the heat flux density in the absence of any particle transport or, in other words, under zero electrical current, we get:

\begin{equation}\label{J=0}
{\bm J}_{N}={\bm 0}=-L_{11}\left( \frac{1}{T}{\bm \nabla }(\mu_{\rm e})\right) +L_{12}{\bm \nabla }(\frac{1}{T})
\end{equation}

\noindent so that the heat flux density under zero electrical current, ${\bm J}_{Q_{J=0}}$, reads:

\begin{equation}\label{Fourier Law}
{\bm J}_{Q_{J=0}}=\frac{1}{T^{2}}\left[ \frac{L_{21}L_{12}-L_{11}L_{22}}{L_{11}}\right] {\bm \nabla}(T)
\end{equation}

\noindent This is the law of Fourier, with the thermal conductivity under zero electrical current given by

\begin{equation}
\kappa_{J}=\frac{1}{T^{2}}\left[ \frac{L_{11}L_{22}-L_{21}L_{12}}{L_{11}}\right]
\end{equation}

We can also define the thermal conductivity $\kappa _{E}$ under zero electrochemical gradient, i.e. under closed circuit conditions:

\begin{equation}
{\bm J}_{Q_{E=0}}=\frac{L_{22}}{T^{2}}{\bm \nabla}(T)=\kappa _{E}{\bm \nabla }(T)
\end{equation}

\noindent It follows that the thermal conductivities $\kappa_{E}$ and $\kappa_{J}$ are simply related through:

\begin{equation}
\kappa_{E}=T\alpha ^{2}\sigma _{T}+\kappa_{J}
\end{equation}

\subsection{Coupled processes}

Let us now shed some light on the coupled processes. In the absence of any electron transport, the basic expression is already known since it is given by Eq. \eqref{J=0}. We may now define the Seebeck coefficient as the ratio between the two forces that derive from the electrochemical and temperature potentials:

\begin{equation}\label{Seebeck coefficient}
\alpha \equiv \frac{-\frac{1}{e}{\bm \nabla }(\mu_{\rm e})}{{\bm \nabla }(T)}=\frac{1}{eT}\frac{L_{12}}{L_{11}}
\end{equation}

Under an isothermal configuration the coupling term between electronic current density and heat flux is obtained from the two following expressions:

\begin{eqnarray}
{\bm J}&=&e{\bm J}_{N}=eL_{11}\left( -\frac{1}{T}{\bm \nabla }(\mu_{\rm e})\right)
\\
{\bm J}_{Q}&=&L_{21}\left( -\frac{1}{T}{\bm \nabla }(\mu_{\rm e})\right)
\end{eqnarray}

\noindent which yield a definition of the Peltier coefficient $\Pi$:

\begin{equation}\label{heat flux}
{\bm J}_{Q}=\frac{1}{e}\frac{L_{12}}{L_{11}}{\bm J} = \Pi {\bm J}
\end{equation}

Now the equality

\begin{equation}\label{Peltier Seebeck coefficient}
\Pi =T\alpha
\end{equation}

\noindent is obvious. The close connexion between Peltier and Seebeck effects is illustrated by this compact expression. In echo to what was said at the begining of Section 3, this shows, from a fundamental point of view, that all thermoelectric effects are in fact different expressions of the same quantity $S_{J}$, called the ``entropy per carrier'' defined by Callen \cite{CallenBook}:
 
\begin{equation}
S_{J}=\alpha e
\end{equation}

\subsection{Kinetic coefficients and general expression for the law of Ohm}

The analysis and calculations developed above allow to establish a \emph{complete} correspondence between the kinetic coefficients and the transport parameters:

\begin{eqnarray}
L_{11}&=&\frac{\sigma _{T}}{e^{2}}T
\\
L_{12}&=&\frac{\sigma _{T}S_{J}T^{2}}{e^{2}}
\\
L_{22}&=&\frac{T^{3}}{e^{2}}\sigma _{T}S_{J}^{2}+T^{2}\kappa _{J}
\end{eqnarray}

\noindent so that the expressions for the electronic and heat currents may take their final forms:

\begin{equation}
{\bm J}_{N}=\frac{\sigma _{T}}{e^{2}}T\left( -\frac{{\bm \nabla }(\mu_{\rm e})}{T}\right) +\frac{\sigma _{T}S_{J}T^{2}}{e^{2}}\left( {\bm \nabla }(\frac{1}{T})\right)
\end{equation}

\begin{equation}
{\bm J}_{Q}=\frac{\sigma _{T}S_{J}}{e^{2}}T^{2}\left( -\frac{{\bm \nabla }(\mu_{\rm e})}{T}\right) +\left[ \frac{T^{3}}{e^{2}}\sigma _{T}S_{J}^{2}+T^{2}\kappa _{J}\right] \left( {\bm \nabla }(\frac{1}{T})\right)
\end{equation}

Since ${\bm J}=e{\bm J}_{N}$, it follows that 

\begin{equation}
{\bm J}=\sigma _{T}{\bm E}-\frac{\sigma _{T}S_{J}}{e}{\bm \nabla }(T)
\end{equation}

\noindent from which we obtain:

\begin{equation}
{\bm E}=\rho _{T}{\bm J}+\alpha {\bm \nabla}(T)
\end{equation}

\noindent where $\rho_T$ is the isothermal conductivity. This is a general expression of the law of Ohm.

\subsection{The dimensionless figure of merit $ZT$}

The off-diagonal terms of the kinetic matrix $\left[{\mathcal L}\right] $ represent the coupling between the heat and the electrical fluxes. The question is now to consider the way to optimize a given material in order to get an efficient heat pump driven by an input electrical current or electrical generators driven by a heat flux. The procedure can be derived for both applications, and we propose here to consider the case of a thermoelectric generator (TEG).

Let us first look at the optimization of the fluxes. Since a thermoelectric system is an energy conversion device, the more heat that flows into the material, the more electrical power may be produced. In order to achieve this, we expect a large thermal conductivity for the material. Unfortunately, this will also lead to a very small temperature difference and consequently a small electrical output voltage and power. This configuration can be called the short-circuit configuration since the fluxes are maximized and
the differences of the potentials minimized.

Let us now consider the coupled processes from the potential point of view. In order to get the larger voltage, the material should experience a large temperature difference between its edges. Then the thermal conductivity of the material should be as small as possible, leading to a very small heat flux and consequently, again, a small electrical output power. This configuration can be called the open-circuit configuration since the potential differences are maximized and the fluxes minimized.

It is worth noticing that both short-circuit and open-circuit configurations lead to an unsatisfactory situation. Moreover they are in
contradiction since the thermal conductivity is expected to be maximum in the short-circuit configuration and minimum in the open-circuit one. The relationship between the thermal conductivities derived above, and written again here: 

\begin{equation*}
\kappa_{E}=T\alpha^{2}\sigma_{T}+\kappa_{J}
\end{equation*}

\noindent offers a way to solve this contradiction. Since it was established under zero current condition, the conductivity $\kappa_{J}$ corresponds to the open-circuit configuration while the conductivity $\kappa_{E}$ corresponds to the short-circuit configuration. From the previous developments we expect the ratio $\kappa_{E}/\kappa_{J}$ to be maximal to get the maximal output electrical power from the TEG. The explicit expression of the ratio contains a definition of the figure of merit $ZT$:

\begin{equation}
\frac{\kappa_{E}}{\kappa_{J}}=\frac{\alpha^{2}\sigma_{T}}{\kappa_{J}}T + 1 = ZT + 1
\end{equation}

A maximal value of the ratio $\kappa_{E}/\kappa_{J}$ implies that reaching the highest possible value of $ZT$ is a necessary condition to qualify a material as optimal. The thermoelectric properties of the material are summarized in the figure of merit $ZT$, as proposed by A. Ioffe \cite{Ioffe}. This quantity gives a quantitative information on the quality of the material, and hence for practical
applications. Since only the material parameters enter into its expression, the figure of merit is clearly the central term for material engineering research. In addition it should be noticed that the present description does not consider at all the anions of the crystal lattice of the thermoelectric material but only the electronic gas\footnote{An analogue situation would be considering a steam engine without any boiling walls.}. This is due to the Onsager description which follows the linear response theory that does not include the lattice contribution to heat conduction. However, this contribution may be accounted for in the general conductance matrix by considering a parallel path for the heat flux.

\section{\label{TEGowc}Device optimization: case of a thermoelectric generator}

\subsection{Device characteristics}
We consider a single-leg module with section $s$ and length $L$ placed between two heat reservoirs that act as thermostats. The proposed thermogenerator configuration includes the presence of heat exchangers, one between the hot thermostat at temperature $T_{\rm hot}$ and the hot side module, the other between the cold side of the module and the cold thermostat at temperature $T_{\rm cold}$. It should be noticed that this configuration cannot be reduced to a simpler one limited to the module, since the hot side, $T_{\rm hM}$, and cold side, $T_{\rm cM}$, module temperatures may vary during the operation of the module.

We assume that the two thermal contacts on both sides of the module are completely characterized by the constant thermal conductances $K_{\rm cold}$ and $K_{\rm hot}$. The total contact conductance, $K_{\rm contact}$, is $K_{\rm contact}^{-1} = K_{\rm cold}^{-1}+K_{\rm hot}^{-1}$. An extension to varying thermal conductances may be considered in the case of non-linear convection or radiative processes. For the sake of simplicity we restrict our analysis to constant heat exchangers thermal conductances. It should also be mentioned that in the case of very efficient heat exchangers, it is tempting to fully neglect their presence and consider a perfect coupling to the thermostats. This consideration is wrong by principle, since, in a dissipative process, a non-zero resistance is always infinitely larger than a truly zero resistance.

\begin{center}
\begin{figure}
\scalebox{.33}{\includegraphics*{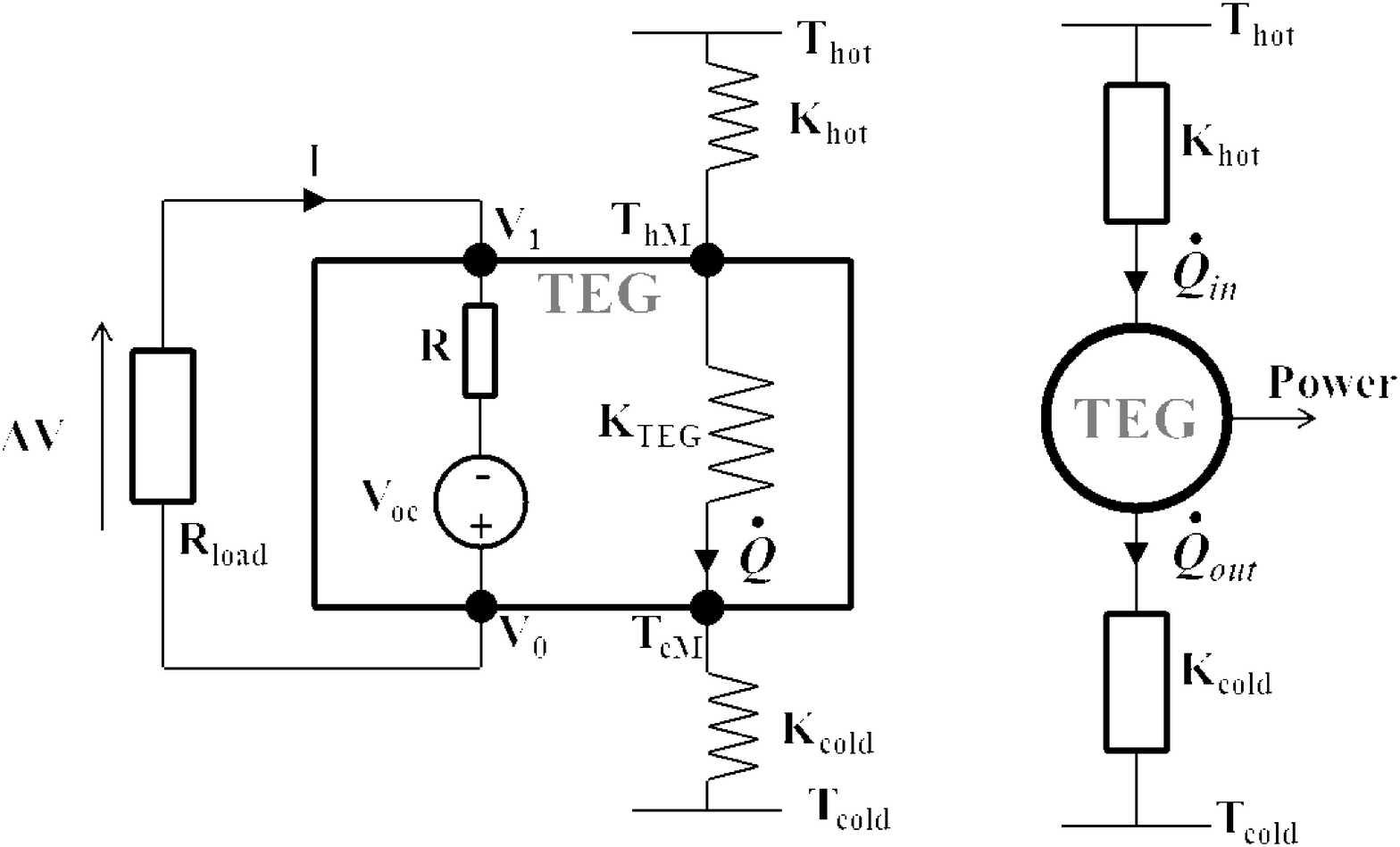}}
\caption{\label{teg} Nodal representation of the thermoelectrical (left) and thermodynamical (right) pictures of the thermoelectric module and the load.}
\end{figure}
\end{center}

The internal electrical resistance is given by $R=L/\sigma_T s.$ The thermal conductance of the module may be given by a current-dependent expression of the type: $K_{\rm TEG}(I)=\kappa(I)s/L$, but we will show below how an expression for the TEG thermal conductance $K_{\rm TEG}$ can be derived using two different ways. As depicted in Fig. \ref{teg}, the open voltage is simply given by $V_{\rm oc}=\alpha \Delta T_{\rm TEG}$ where $\Delta T_{\rm TEG} = T_{\rm hM} - T_{\rm cM}$, which is some fraction of the total temperature difference $\Delta T = T_{\rm hot}-T_{\rm cold}$.  The TEG is also characterized by its isothermal electrical resistance $R$, and the Seebeck coefficient $\alpha$. The thermal conductance $K_{\rm TEG}$ reduces to the conductance $K_{_{V=0}}$, under zero voltage (electrical short circuit), and to the conductance $K_{_{I=0}}$, at zero electrical current (open circuit). Note that both electrons and phonons contribute to the thermal conductance. The figure of merit $ZT$ is given by : $ZT = \alpha^2T/RK_{_{I=0}}$.

\subsection{Thermal and electrical currents}
For a temperature difference across the TEG, $\Delta T_{\rm TEG}$, that is not too large, it is safe to assume that the incoming and outgoing heat fluxes are linear in the temperature difference $\Delta T_{\rm TEG}$. This permits a description of the TEG characteristics with the force-flux formalism, which yields the following equations:

\begin{equation}\label{frcflx}
\left(
\begin{array}{c}
I\\
I_Q\\
\end{array}
\right)
=
\frac{1}{R}~
\left(
\begin{array}{cc}
1~ & ~\alpha\\
\alpha {\overline{T}}~ & ~\alpha {\overline{T}}^2 + RK_{_{I=0}}\\
\end{array}
\right)
\left(
\begin{array}{c}
\Delta V\\
\Delta T_{\rm TEG}\\
\end{array}
\right),
\end{equation}

\noindent where $I$ is the delievered electrical current that flows through the load, which is assumed to be simply resistive, and $I_Q$ is the thermal current through the TEG; $\Delta V$ is the voltage across the TEG. The open-circuit voltage is $V_{\rm oc} = \alpha\Delta T_{\rm TEG}$. The average temperature in the module is simply taken as $\overline{T} = (T_{\rm cM}+T_{\rm hM})/2$.

The thermal current is the sum of the contributions of convective heat transfer, i.e. heat transported within the electrical current, and steady-state conduction \cite{Apertet2012e}, usually associated with Fourier's law:

\begin{equation}\label{IQ1}
I_Q = \alpha \overline{T} I + K_{_{I=0}} \Delta T_{\rm TEG}
\end{equation}

\noindent Ohm's law applies as follows: $\Delta V = -R_{\rm load} I$ and the electrical current $I$ reads:

\begin{equation}\label{Int1}
I = \frac{\Delta V + \alpha \Delta T_{\rm TEG}}{R} = \frac{\alpha \Delta T_{\rm TEG}}{R_{\rm load} + R}
\end{equation}

\noindent After substitution of the electrical current $I$ into Eq.~(\ref{IQ1}), the TEG thermal conductance $K_{\rm TEG}$ can be defined with:

\begin{equation}\label{IQ2}
I_Q = \left( \frac{\alpha^2 \overline{T}}{R_{\rm load} + R} +K_{_{I=0}} \right)\Delta T_{\rm TEG} = K_{\rm TEG}\Delta T_{\rm TEG}
\end{equation}

\noindent It is important to note that the dependence of the thermal conductivity $K_{\rm TEG}$ on the electrical operating point since its expression contains the load electrical resistance. It is useful at this point to obtain an expression of $K_{\rm TEG}$ in a different fashion. We start from the relationship between the two thermal conductances $K_{_{V=0}}$ and $K_{_{I=0}}$ of the TEG \cite{goupil2}:

\begin{equation}\label{thermconds}
K_{_{V=0}} = K_{_{I=0}} \left( 1+ Z\overline{T}\right)
\end{equation}

\noindent which may be extended to the following formula:
\begin{equation}\label{kteg2}
K_{\rm TEG}(I) = K_{_{I=0}} \left( 1+ \frac{I}{I_{\rm sc}} Z\overline{T}\right),
\end{equation}

\noindent where we introduced the short circuit current $I_{\rm sc} = \alpha\Delta T_{\rm TEG}/R$ such that $K_{\rm TEG}(I_{\rm sc}) = K_{_{V=0}}$. Equations (\ref{kteg2}) and (\ref{IQ2}) yield exactly the same expression for $K_{\rm TEG}$. It is important to see that since the short circuit current $I_{\rm sc}$ does depend on the effective temperature difference across the TEG, $\Delta T_{\rm TEG}$, there is no closed form solution for the global distributions of electrical and heat currents and potentials in the device.

Figure~\ref{teg} shows that the electrical part of the TEG may be viewed as the association of a perfect generator and a resistance which is the physical resistance of the generator. By definition, the open circuit voltage of the TEG depends on the temperature difference across the TEG. Because of the presence of finite thermal contact conductances $\Delta T_{\rm TEG}$ depends on the electrical load and $V_{\rm oc}$ cannot be considered as the output voltage of a perfect Th\'evenin generator since its characteristics depend on the load:

\begin{equation}\label{voc}
V_{\rm oc} =\alpha \Delta T \frac{K_{\rm contact}}{K_{{_{I=0}}}+K_{\rm contact}}-I R \frac{Z\overline{T}}{1+K_{\rm contact}/K_{_{I=0}}}
\equiv V_{\rm oc}' - IR'
\end{equation} 

\noindent The first term on the right hand side is independent of the electrical load, the second depends on the electrical current delivered. Now, the open-circuit voltage $V_{\rm oc}'$ and the internal resistance is $R_{\rm TEG}=R+R'$ thus introduced permit the rigorous definition of a perfect Th\'evenin generator.

\subsection{Computation of the temperature difference across the TEG}
The analysis developed so far assumes an explicit knowledge of $\Delta T_{\rm TEG}$, but for practical and modeling purposes it is more useful to obtain expressions of power and efficiency to be optimized as functions of the temperature difference between the two reservoirs, $\Delta T = T_{\rm hot} - T_{\rm cold}$. To determine $\Delta T$, the easiest option is to start from Ioffe's approach \cite{Ioffe} to the definitions of the incoming heat flux $\dot{Q}_{\rm in}$ and outgoing heat flux $\dot{Q}_{\rm out}$:

\begin{eqnarray}
\label{Qin}
\dot{Q}_{\rm in} &=& \alpha T_{\rm hM}I - \frac{1}{2} RI^2 + K_{_{I=0}}(T_{\rm hM} - T_{\rm cM})\\
\label{Qout}
\dot{Q}_{\rm out} &=& \alpha T_{\rm cM}I + \frac{1}{2} RI^2 + K_{_{I=0}}(T_{\rm hM} - T_{\rm cM})
\end{eqnarray}

\noindent Since we also have $\dot{Q}_{\rm in} = K_{\rm hot}(T_{\rm hot} - T_{\rm hM})$ and $\dot{Q}_{\rm out} = K_{\rm cold}(T_{\rm cM} - T_{\rm cold})$, these equations define a 2$\times$2 system which links $T_{\rm cM}$ and $T_{\rm hM}$ to $T_{\rm cold}$ and $T_{\rm hot}$:

\begin{equation}\label{temps}
\left(
\begin{array}{c}
\phantom{-}T_{\rm hot~} + \frac{\displaystyle 1}{\displaystyle 2}\frac{\displaystyle RI^2}{\displaystyle K_{\rm hot~}}\\
~\\
-T_{\rm cold} - \frac{\displaystyle 1}{\displaystyle 2}\frac{\displaystyle RI^2}{\displaystyle K_{\rm cold}}\\
\end{array}
\right)
=
\left(
\begin{array}{cc}
{\mathcal M}_{11} & {\mathcal M}_{12}\\
~\\
{\mathcal M}_{21} & {\mathcal M}_{22}\\
\end{array}
\right)
\left(
\begin{array}{c}
T_{\rm hM}\\
~\\
T_{\rm cM}\\
\end{array}
\right),
\end{equation}

\noindent where the four dimensionless matrix elements are given by:
${\mathcal M}_{11} =  K_{_{I=0}}/K_{\rm hot} + \alpha I/K_{\rm hot} +1$, 
${\mathcal M}_{12} =  - K_{_{I=0}}/K_{\rm hot}$, 
${\mathcal M}_{21} =  K_{_{I=0}}/K_{\rm cold}$, 
${\mathcal M}_{22} =  \alpha I/K_{\rm cold} - K_{_{I=0}}/K_{\rm cold} - 1$.

\noindent The analytic expressions of the temperatures $T_{\rm hM}$ and $T_{\rm cM}$ are easily obtained by matrix inversion, but the exact expression of $\Delta T_{\rm TEG}$ as a function of $T_{\rm hot}$ and $T_{\rm cold}$ is cumbersome, and truly not necessary for the discussions that follow in the Chapter. However, an approximate but straightforward relationship between $\Delta T_{\rm TEG}$ and $\Delta T$ certainly is worthwhile. Using an analogue of the voltage divider formula, this relationship may be obtained by assuming that the thermal flux is constant in the whole system:

\begin{equation}\label{kapctc}
\Delta T_{\rm TEG} = T_{\rm hM} - T_{\rm cM} \approx \frac{K_{\rm contact}}{K_{\rm TEG}+K_{\rm contact}}~\Delta T
\end{equation}

\section{Maximization of power and efficiency with fixed $Z\overline{T}$}

\subsection{Maximization of power by electrical impedance matching}

The electrical power produced by the TEG can be simply expressed as

\begin{equation}\label{Pprod1}
P = \frac{{V_{\rm oc}'}^2 R_{\rm load}}{(R_{\rm TEG}+R_{\rm load})^2},
\end{equation}

\noindent which shows that maximization of the produced output power for a given thermal configuration is achieved when

\begin{equation}\label{adaptelec}
R_{\rm load}=R_{\rm TEG}
\end{equation}

\noindent Expressed in a more traditional way using the ratio $m = R_{\rm load}/R$ defined by Ioffe \cite{Ioffe}, the condition for maximization reads:

\begin{equation}\label{rload}
m_{_{P=P_{\rm max}}} = 1 + \frac{Z\overline{T}}{K_{\rm contact}/K_{_{I=0}}+1},
\end{equation}

\noindent Because of the presence of an additional part in the equivalent resistance $R_{\rm TEG}$ of the TEG due to the finite thermal contact coupling, the electrical impedance matching condition (\ref{rload}) does not correspond to the condition $m=1$ (or, equivalently, $R_{\rm load}=R$) of the ideal case. This was seen previously by Freunek and co-workers \cite{Freunek}.

When the electrical resistance matching is achieved, the maximum ouput power reads:

\begin{equation}\label{pmax}
P_{\rm max} = \frac{(K_{\rm contact}\Delta T)^2}{4(K_{_{I=0}}+K_{\rm contact})\overline{T}}~\frac{Z\overline{T}}{1+Z\overline{T}+K_{\rm contact}/K_{_{I=0}}},
\end{equation}

\subsection{Maximization of power by thermal impedance matching}

The choosing of the thermal properties of the TEG so that a maximum output power is obtained directly relates to the general problem of the optimization of the working conditions of a non-endoreversible engine coupled to the temperature reservoirs through finite conductance heat exchangers. For endoreversible engines the heat exchangers are the only location for entropy production, a process thus governed by only one degree of freedom. For non-endoreversible engines, entropy is produced inside the engine because of the Joule effect and the thermal conduction effect; so this confers two additional degrees of freedom to the system. The framework presented in this Chapter thus extends the classical so-called Novikov-Curzon-Ahlborn configuration \cite{Chambadal,Novikov,curzahl,bejan96} specialized to endoreversible engines.

If the total conductance $K_{\rm contact}$ is fixed by an external constraint, optimization of power may be achieved with respect to $K_{_{I=0}}$ through the following condition:

\begin{equation}\label{thermaladapt}
\frac{K_{\rm contact}}{K_{_{I=0}}} = 1 + \frac{Z\overline{T}}{1+m},
\end{equation}

\noindent which corresponds to the satisfaction of the equality: 

\begin{equation}\label{kctcteg}
K_{\rm contact} = K_{\rm TEG}
\end{equation}

\noindent The above equation is similar to that derived by Stevens in Ref.~\cite{Stevens2001} who saw that the thermal impedance matching corresponds to the equality between the thermal contact resistance and that of the TEG. The difference with our result, Eq. \eqref{kctcteg}, is that the thermal resistance used in Ref.~\cite{Stevens2001} for the thermoelectric module is obtained under open circuit condition and hence does not account for the convective part of the thermal current, while $K_{\rm TEG}$ defined in Eq.~(\ref{IQ2}) does. To end this section, we emphasize the similarity between electrical and thermal impedance matchings respectively given by equations (\ref{rload}) and (\ref{thermaladapt}).

\subsection{Simultaneous thermal and electrical impedance matching}

For a particular configuration imposed by the environnement, the optimal point for the operation of a TEG may be found by joint optimization of the electrical and thermal conditions, which we do by solving equations (\ref{rload}) and (\ref{thermaladapt}) \emph{simultaneously}. We find:

\begin{eqnarray}\label{k0ZT}
\frac{K_{\rm contact}}{K_{_{I=0}}} & = & \sqrt{Z\overline{T}+1}
\\
\label{mZT}
m_{_{P=P_{\rm max}}} & = & \sqrt{Z\overline{T}+1}
\end{eqnarray}

\noindent Note that Eq.~(\ref{k0ZT}) was presented by Freunek and co-workers \cite{Freunek}, and that Yazawa and Shakouri obtained both equations \cite{Yasawa}. With these two impedance matching conditions, we find that the maximum power produced by the TEG is given by:

\begin{equation}
P_{\rm max} = \frac{K_{\rm contact} Z\overline{T}}{\left(1+\sqrt{1+Z\overline{T}}\right)^2} \frac{(\Delta T)^2}{4\overline{T}},
\end{equation}

\subsection{On the importance of thermal impedance matching}

\begin{center}
\begin{figure}
\scalebox{.3}{\includegraphics*{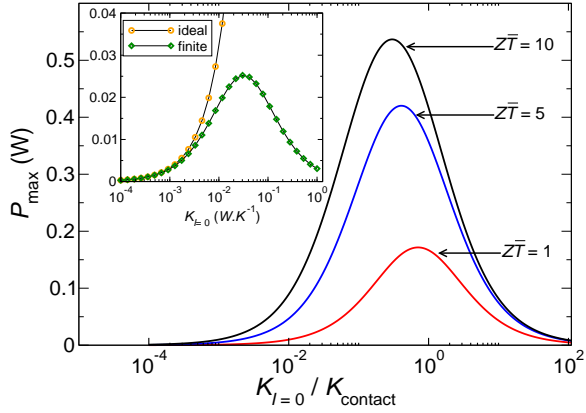}}
\caption{\label{fig2} Maximum power as a function of the ratio $K_{_{I=0}}/K_{\rm contact}$ for various $Z\overline{T}$ values at fixed $K_{\rm contact}$ (notice the use of a logarithmic scale for the abscissa axis). In the inset, the curves (with ideal and finite thermal contacts) are computed with the data of Ref.~\cite{nemirbeck} where the authors studied $P_{\rm max}$ for three values of $K_{_{I=0}} : 3\times 10^{-3},~ 6\times 10^{-3}$ and $1.2\times 10^{-2}$ W$\cdot$K$^{-1}$.}
\end{figure}
\end{center}

\noindent The variations of the maximum power $P_{\rm max}$ as a function of the ratio $K_{_{I=0}}/K_{\rm contact}$ [Eq.~(\ref{pmax})] are shown in Fig.~\ref{fig2} for three values of the figure of merit $Z\overline{T}$. Though, as expected, higher values of $Z\overline{T}$ yield greater values for the maximum of $P_{\rm max}$ and larger widths at half maximum, the curves displayed in Fig.~\ref{fig2} highlight the importance of thermal impedance matching: a high value of $Z\overline{T}$ does not guarantee a greater $P_{\rm max}$ for \emph{any} value of the thermal conductance at zero electrical current $K_{_{I=0}}$; for instance, $P_{\rm max}$ at $K_{_{I=0}}=K_{\rm contact}$ for $ZT=1$ is greater than $P_{\rm max}$ at $K_{_{I=0}}= 5K_{\rm contact}$ for $ZT=10$.

In the inset of Fig.~\ref{fig2}, two curves represent the maximum power as a function of $K_{_{I=0}}$ for \emph{finite} and perfect thermal contacts respectively; these permit an understanding of why the TEG with the highest $K_{_{I=0}}$ presents the largest performance degradation, a fact that was observed by Nemir and Beck \cite{nemirbeck}: to analyze the impact of thermal contacts on device performance, they considered various configurations giving the same value for the figure of merit $Z{\overline T}$. They found that for a given value of contact thermal conductance, the device performance is strongly influenced by how the fixed figure of merit of the thermoelectric module is achieved. 

\subsection{Maximum efficiency}
We now turn to the efficiency that characterize the conversion of the heat current $I_{Q}$ into the electric power $P$: $\eta = P/I_{Q}$. An explicit expression for $\eta$ is

\begin{equation}\label{etag}
\eta = \frac{K_{\rm contact}+K_{\rm TEG}}{K_{\rm contact}K_{\rm TEG}}\frac{P}{\Delta T},
\end{equation}

\noindent considering Eqs. (\ref{IQ2}) and (\ref{kapctc}). This expression reduces to
\begin{equation}
\eta = \eta_{\rm C} \times \frac{m}{1 + m + \left( ZT_{\rm hot}\right)^{-1}\left( 1+ m \right)^2 - \eta_{\rm C}/2}
\end{equation}

\noindent in the case of ideal thermal contacts.

The value of $m$ which maximizes the efficiency (\ref{etag}) is:

\begin{equation}\label{metamax}
m_{_{\eta=\eta_{\rm max}}} = \sqrt{\left(1+Z\overline{T}\right)\left(1+Z\overline{T} \frac{K_{_{I=0}}}{K_{\rm contact}+K_{_{I=0}}}\right)}
\end{equation}

\noindent It explicitly depends on the thermal conductances  $K_{\rm contact}$ and $K_{_{I=0}}$.

\subsection{Analysis of optimization and power-efficiency trade-off}

\begin{figure}
\scalebox{.3}{\includegraphics*{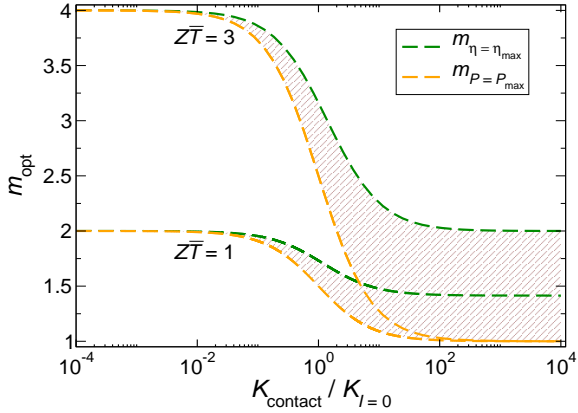}}
\caption{\label{fig3}Variations of the optimal parameters $m_{_{\eta=\eta_{\rm max}}}$ (dashed line) and $m_{_{P=P_{\rm max}}}$ (dashed-dotted line) as functions of $K_{\rm contact}$ scaled to $K_{_{I=0}}$, for $Z\overline{T}=1$ and $Z\overline{T}=3$. The shaded areas corresponds to the region of best optimization.}
\end{figure}

If, as for liquid-gas heat exchangers, the working conditions lead to modifications of $K_{\rm contact}$, the operating point of the thermoelectric device necessarily changes. It is then worth checking whether $m_{_{\eta=\eta_{\rm max}}}$ is bounded or not when the ratio $K_{_{I=0}}/K_{\rm contact}$ varies. The mean temperature $\overline{T}$ varies very little with $K_{\rm contact}$ so we may safely consider that the figure of merit is fixed without loss of generality for the discussion that follows.

We saw that the optimal values of the electrical load to achieve maximum power or efficiency are different from those of the ideal case when finite thermal contacts are accounted for in the TEG model. Considering two values of the figure of merit: $Z\overline{T}=1$ and $Z\overline{T}=3$, the optimal parameters $m_{\rm opt}$ [$m_{_{\eta=\eta_{\rm max}}}$ for maximum efficiency in Eq.~(\ref{metamax}), and $m_{_{P=P_{\rm max}}}$ for maximum power, in Eq.~(\ref{rload})] are plotted against $K_{\rm contact}$ (scaled to $K_{_{I=0}}$) in Fig. \ref{fig3}. For a given value of $Z\overline{T}$ and conditions close to perfect thermal contacts, i.e. $K_{\rm contact} \gg K_{_{I=0}}$, the maximum power and maximum efficiency are well separated: $m_{_{\eta=\eta_{\rm max}}} \longrightarrow \sqrt{1+Z\overline{T}}$, and $m_{_{P=P_{\rm max}}} \longrightarrow 1$. Note that the separation between both can only increase with $Z\overline{T}$.

In the opposite limit, i.e. for $K_{\rm contact} \ll K_{_{I=0}}$ we see that $m_{_{\eta=\eta_{\rm max}}} \longrightarrow 1+Z\overline{T}$, which is the upper bound to $m_{_{P=P_{\rm max}}}$ too: both optimal parameters coincide. The convergence of these two optimal parameters towards the same value could be seen as interesting in the sense that this fact implies that there is no need for a trade-off between efficiency and power; however this also implies a significant performance decrease. As a matter of fact, the optimal regions to satisfy the power-efficiency trade-off are those lying between each pair of curves, and we see that the narrowing of these zones, which also comes along with lower values of $Z\overline{T}$, is undesirable since, consequently, less flexibility in terms of working conditions of the thermoelectric generator is allowed.

\begin{figure}
\scalebox{.3}{\includegraphics*{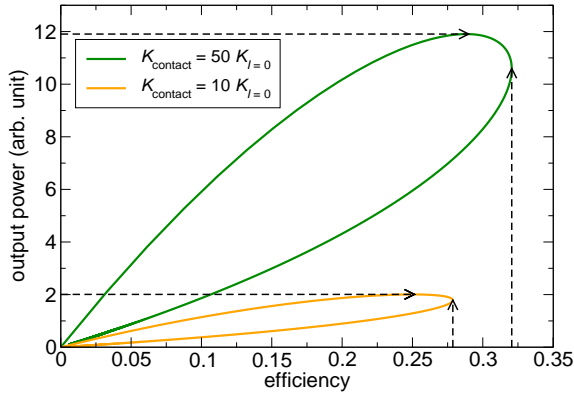}}
\caption{\label{fig4}Power versus efficiency curves for two cases with a fixed figure of merit $Z\overline{T} = 1$. Only the value of $K_{\rm contact}$ varies.}
\end{figure}

The figure \ref{fig4} displays two power-efficiency curves, one for $K_{\rm contact} = 10 K_{_{I=0}}$, the other for $K_{\rm contact} = 50 K_{_{I=0}}$. As the contact thermal conductance decreases, we observe a narrowing of the optimal zone in accordance with our analysis of Fig. \ref{fig3}. The arrows indicate the maximal values that $P_{\rm max}$ and $\eta_{\rm max}$ may take. As the quality of the thermal contacts deteriorates the power-efficiency curve reduces to a point located at the origin.

\section{Summary and discussion on efficiency at maximum power}
We presented the force-flux formalism developed by Onsager \cite{onsager1, onsager2} and later adapted to thermoelectricity by Callen \cite{callen1} and Domenicali \cite{Domenicali}. This formalism provides very convenient tools to obtain and analyze in simple terms the thermal and electrical conditions which allow the maximum power production by a thermoelectric generator \emph{non-ideally} coupled to heat reservoirs. These conditions are expressed very simply for the thermal impedance matching: the equality of the contact thermal conductances and the equivalent thermal conductance of the TEG. It is crucial to see that the quality of the contacts between the TEG and the heat reservoirs is important since high values of $Z\overline{T}$ are of limited interest otherwise. We also saw that the nontrivial interplay between the thermal and electrical properties of the TEG makes difficult the search for the optimum conditions for maximum output power production since both electrical and thermal impedance matching must be satisfied simultaneously. Note that this is also the basic idea underlying the compatibility approach \cite{snyder2003} for ideal thermoelectric systems.

Searching for ways to increase the efficiency of thermoelectric conversion processes in devices operating in various environmental conditions boosts a variety of research activities spanning from materials sciences to nonequilibrium statistical mechanics at the mesoscale, and device modeling. All these areas of research are highly topical. Approaches based on the concept of minimization of entropy production have been developed and proved succesful in various sectors of thermal engineering and sciences \cite{bejan96}; new contributions continue to feed this field with, e.g., the recent introduction of the thermoelectric potential \cite{goupil2,snyder2003,goupil1}. Since one is interested in power rather than energy, thermoelectric efficiency has been analyzed in the frame of finite-time thermodynamics \cite{FTT1, FTT2, FTT3, Andresen2011}, focusing on the efficiency at maximum power \cite{Apertet2012c, Apertet2012d}. Such efficiency was initially shown to be bounded by the so-called formula of Curzon and Ahlborn (CA) in the specific case of endoreversible heat engines \cite{curzahl}. The analysis of Curzon and Ahlborn was later put on firmer grounds in the frame of linear irreversible thermodynamics assuming a strong coupling between the heat flux and the work \cite{vandenbroeck} and further discussed by Apertet and co-workers \cite{Apertet2012c, Apertet2012d}. Analyses of efficiency at maximum power for nanoscopic quantum dot systems \cite{esposito} and extension to stochastic heat engines \cite{schmiedl} are also being developed.

The work presented in this Chapter and in research papers focusing on thermoelectric devices \cite{Apertet2012c, Apertet2012d, Apertet2012a, Apertet2012b} significantly advance the understanding of the central concept of energy conversion efficiency of heat engines. Since the advent of finite time thermodynamics, the Curzon and Ahlborn efficiency has become a paradigm. Given the utmost importance of the physics of heat engines, which are ubiquitous in the sense that they are systems useful to model biological cells, mesoscopic solid-state systems, macroscopic devices, we think that the CA efficiency, though well established through the years, deserves close inspection. A very important work was initiated by Schmiedl and Seifert (SS), but the crucial question of the discrepancy between their result and the CA efficiency remained without answers until now. The answer, far from trivial, required an in-depth analysis on the nature of irreversibilities in heat engines \cite{Apertet2012c, Apertet2012d}. If one overlooks some of the subtleties that we put forth, one may not succeed in providing a sound and clear framework that explains in a transparent fashion the energy conversion efficiency of heat engines. Our work, to some extent, provides a shift in paradigm: we have come not only to propose clear definitions of both CA and SS efficiencies but also show that a system may undergo a continuous transition from one type of efficiency to the other by tuning the different sources of irreversibility. These facts are crucial for any optimization work.

\section{\label{meso}Outlook on the next frontier: the mesoscopic scale}
Mesoscopic systems offer an interesting field of play at the crossroads of quantum and classical physics, to experimentalists as well as theoreticians. Indeed, the study of mesoscopic systems cannot be reduced to practical work aiming only at miniaturization of transistors for microelectronics applications. From a fundamental viewpoint, it is in these systems that the size of fluctuations become sufficiently important as compared to averages, so that their influence on the systems' properties emerges. The onset of quantum effects at the mesoscopic scale depends on the size of the considered system, its temperature and external constraints that depend on the interactions with its environment.

Over the last 30 years great strides were made in fabrication of artificial structures down to very small scales. Operation of a number of modern devices rely on a proper understanding of the physics of nano- and mesoscale structures such as, e.g., semiconductor quantum wells, superlattices, and quantum dots, which today are routinely produced as high quality custom-made samples. Mesoscopic systems are characterized by a great number of constituents. For some applications or fundamental studies involving conversion of transfer of energy, they can be considered as heat engines; so it is tempting to describe and analyze them using the concepts and terminology that derive from classical thermodynamics. As a matter of fact, such notions as, e.g., entropy and entropy production, which we saw in Section 2, must be revisited considering that mesoscopic devices operate in regimes far from equilibrium where fluctuations are strong. This implies that transport phenonema and the related measurable quantities in these systems must be identified and understood properly.

The experimental study of electron transport, which is typically ballistic and coherent in mesoscopic systems, can be performed with the so-called quantum point contacts (QPCs); these narrow, confining constrictions are made up between wider conducting regions of the system under consideration, and their width is comparable to the electrons' wavelength at low temperatures. The quantization of ballistic electron transport through such constriction demonstrates that conduction \emph{is} transmission. Because of conductance quantization \cite{Wees} in quantum point contacts, the Seebeck coefficient was analyzed at the  threshold energies of the conductance plateaus, where the change in the conductance is very important. Van Houten and co-workers obtained the Seebeck coefficient in a system based on QPC in a two dimentional electronic gas of a GaAs-AlGaAs semiconducting heterostructures \cite{VanHouten}. They concluded that the thermopower exhibits quantum size effects and oscillates each time a new mode opens up in the QPC.

For thermoelectric systems at this scale, this calls for the development of recent approaches such as, e.g., finite time thermodynamics \cite{FTT1, FTT2, FTT3, Andresen2011} and stochastic thermodynamics \cite{VanDenBroeck86,Seifert08,Sekimoto10,Esposito12}, and their association to those which proved fruitful for the computation and measurement of the thermoelectric transport coefficients. The Landauer-Buttiker formalism provides necessary tools to consider nano-systems placed between several reservoirs and study the multichannel scattering. Sivain and Imry used this approach to compute and study linear transport coefficients of a thermoelectric sample characterized by some disorder in a case where the connections to the chemical and temperature reservoirs are achieved with ideal multichannel leads \cite{Imry}. Dissipation processes due to inelastic scattering were assumed to occur only in the reservoirs. By looking at the thermopower near the mobility edge, they pointed out some deviations of the kinetic coefficients from Onsager's relations and the Seebeck coefficient from the Cutler-Mott formula.

More recent works \cite{Humphrey2005,Nakpathomkun2010,Karlstrom2011} challenge the view that thermoelectric heat engines are by nature irreversible in their operation. The purpose of these indeed is to find ways to allow reversible diffusive transport in thermoelectric materials and the proposed route is that of nanostructures, which if they are sufficiently well tailored, permit the narrowing of the charge carriers' densities of states (DOS). The idea is that if electron transport is limited to a narrow energy band which corresponds to an energy such that the two Fermi functions characterizing the hot and cold reservoirs respectively, are equal, then together with a weak electron-phonon, coupling, friction effects are drastically reduced. While delta-function may represent an ideal limit, quantum confinement effects generate a finite lower bound to the DOS widths, and hence limit the efficiency of the device.

Our view on the matter of irreversibility derives from the main message of finite time thermodynamics: trading a part of the efficiency for the ability to produce power is possible only if irreversible processes are introduced in the thermodynamic cycle. Further, using numerical simulations we demonstrated, in the case of a two thermally coupled macroscopic heat engines, that efficiency at maximum power is increased when the hot-side Joule heating is favoured \cite{Apertet2012d}. We interpreted this result as a recycling of the degraded energy: if it is evacuated to the hot heat reservoir, this energy becomes available to be used again, while if it is evacuated to the cold side, it is irretrievably lost. These considerations must be examined at the mesoscale, where irreversible thermodynamics has not completely given way to reversible dynamics. Finally, we showed that optimization of the operation of a TEG must simultaneously satisfy electrical and thermal impedance matching. At the mesoscale the notion of impedance matching must be considered with care. Adapation of our analysis to mesoscale thermal engines is underway, and we have every reason to be optimistic.

\end{document}